\documentclass{appolb}
\usepackage{graphicx}
\usepackage{amssymb}

\def\ran{\rangle}

\begin{document}
\title{Spectroscopy with glueballs and the role of
$f_0(1370)$
\thanks{Presented at workshop ``Excited QCD 2013'', Bjelasnica Mountain, Sarajevo,
Bosnia-Herzegovina, Feb. 3-9, 2013.}%
}
\author{Wolfgang Ochs
\address{Max-Planck-Institut f\"ur Physik, F\"ohringer Ring 6, D-80805 M\"unchen,
Germany}
}
\maketitle
\begin{abstract}
The existence of glueballs, bound states of gluons, is one of the basic
predictions of QCD; the lightest state is expected to be a scalar. 
The experimental situation, however, is still ambiguous. 
The existence of $f_0(1370)$ would point to a supernumerous state within the
nonet classification of scalars and would therefore provide a hint towards a
glueball.
In this talk we summarise some arguments in favour and against
the existence of $f_0(1370)$ and discuss schemes with and without this state
included.  
\end{abstract}
  
\section{Introduction}
The existence of glueballs is a consequence of the self-interaction of
gluons in QCD with consequences studied
already about 40 years ago
 \cite{Fritzsch:1975tx}. Today, within the lattice QCD 
approach,  
the mass of the lightest scalar glueball is found around 1700 MeV in the
theory with gluons only 
while in the full theory the mass is found to drop to
$\sim 1000$ MeV \cite{Hart:2006ps} or to stay largely unchanged
\cite{Richards:2010ck,Gregory:2012hu}.
QCD sum rules predict scalar gluonic mesons 
as well in the range 1000-1700 MeV
\cite{Narison:2005wc,Forkel:2003mk,Harnett:2008cw}.
The experimental search for a scalar glueball has lead to several scenarios
for a spectroscopy with glueballs.
A recent status of theoretical 
and experimental results on glueballs is found in
\cite{Ochs:2013gi}.

One strategy to find glueballs is based on the identification of the
scalar nonets lowest in mass 
and the search for super-numerous isoscalar states 
which could be related to the 
presence of a glueball 
in the spectrum. In addition, pure glueballs are
characterised by flavour symmetric decays 
(with possible modifications \cite{Ochs:2013gi}) and they are
expected to be predominantly produced in ``gluon-rich'' processes.
In general, glueballs could mix with isoscalar quarkonium states. 
An important example for such a supernumerous state is $f_0(1370)$ which we
will discuss here in particular.

\section{Scalar meson spectrum with $f_0(1370)$}

The states above 1 GeV listed by the Particle Data Group  
\cite{beringer2012pdg} 
$$ f_0(1370),\ f_0(1500),\ f_0(1710),\ K^*_0(1430),\ a_0(1450),$$
can be interpreted as being formed by a nonet of $q\bar q$ states 
and a glueball where the two $q\bar q$ isoscalars
and the glueball mix into three isoscalar $f_0$'s.
Such a scheme has been suggested originally by Amsler and Close
\cite{Amsler:1995td}. At that time the newly discovered $f_0(1500)$ meson
has been related to the glueball with mass predicted near 1500 MeV
by lattice theory with gluons only.
The closer inspection of the decay
branching ratios, however, suggested a mixing scheme for the 
three $f_0$ mesons. 
Other mixing schemes for the glueball are considered in
\cite{Giacosa:2005zt,Cheng:2006hu,Albaladejo:2008qa}, for a review, see
\cite{Crede:2008vw}. 

The states below 1 GeV 
$$ f_0(500)/\sigma,\ f_0(980),\ K^*(900)/\kappa,\ a_0(980)$$
can be grouped into a light meson nonet\footnote{The $K^*(900)/\kappa $
is not considered as established by the PDG at present.}
formed by $q\bar q$ (as in
\cite{Morgan:1974cm,Tornqvist:1995ay,vanBeveren:1998qe}) or by diquark
bound states
(as in \cite{Jaffe:1976ig,Achasov:1987ts,Hooft:2008we}). 

\section{Evidence for $f_0(1370)$ revisited}
The crucial element in these schemes with glueball is the existence of
$f_0(1370)$ and therefore we will reconsider the evidence. In the actual
edition of the PDG the rather
wide ranges for mass and width are reported 
\begin{equation}
M=1200-1500\ {\rm MeV},\ \Gamma= 200 - 500\ {\rm MeV}.
\end{equation}
There are 12 decay channels ``seen'':
$\pi\pi,\ K\bar K,\ \eta\eta,\ 4\pi, \ 
\gamma\gamma $ and various sub-channels of $4\pi$,
but no single experimental number on branching ratios nor ratios thereof has
been determined because of conflicting results. This is quite different from
the nearby $f_0(1500)$ with five well established branching ratios. 
Accordingly,
supportive  \cite{Bugg:2007ja} and sceptical views \cite{Klempt:2007cp} 
about $f_0(1370)$ have been presented in the past. A detailed discussion
of various observations is given in \cite{Ochs:2013gi}. Here we present an
overview and some details of two energy independent analyses.

\subsection{Overview}

The evidence for $f_0(1370)$ has been presented 
first in  $p\bar p$ annihilation at rest by the
Crystal Barrel Collaboration (CBAR) 
\cite{Amsler:1995bf}  
and this state 
has been studied together with
$f_0(1500)$ in the reactions
$a)\ p\bar p\to \pi^0\pi^0\pi^0,\quad b)\ p\bar p\to \pi^0\eta\eta,\quad
c)\ p\bar p\to \pi^0\pi^0\eta.$ 
Signal bands in the Dalitz plots related to $f_0(1500)$ are always clearly
visible. A signal from $f_0(1370)$ can be seen in the $\eta\eta$ channel
but there is an interference with the $\eta\pi$ resonances in
crossed channels. The $f_0(1370)$ signal disappears immediatly 
if the cms energy is
increased above the $p\bar p$ threshold. 
So the evidence for $f_0(1370)$ relies on the proper global
multi-channel fit with various interfering resonances present.

These problems are avoided if the $S$ wave amplitudes are reconstructed
in an energy independent analysis in a sequence of mass bins with
sufficiently high statistics. The resonant
behaviour is then found from the characteristic behaviour of the 
complex amplitude. Such results are available for 2-body
$\pi\pi\to ab$ scattering, which can be reconstructed from 
$\pi p\to ab+n$ processes. Proper care has to be taken in these analyses 
in the treatment of nucleon spins.
Other high statistics results became available recently from
$D\to 3\pi$ where the spin problems disappear or from $B\to J/\psi\pi\pi$
with only one spinning particle.
In central production processes $pp\to p+X+p$ the measurements
at SPS energies are difficult to analyse because of non-trivial
superpositions from processes with different nucleon helicities
\cite{Ochs:2013gi}. 

A large fraction of $f_0(1370)$ decays goes into $4\pi$ channels 
($\gtrsim 70\%)$. Here
different experiments on central production
and $p\bar p$ annihilation 
provide conflicting results; furthermore no evidence for the
existence of two resonances at 1370 and 1500 MeV has been found 
\cite{Klempt:2007cp}. 
Here some clarification is necessary.
In this note we restrict ourselves to some 2-body processes where
energy independent phase shifts are available.

\subsection{Search in phase shift analysis of $\pi\pi$ scattering}
Such data are extracted from the reaction $\pi p\to \pi \pi n (\Delta)$ 
in application of the
one-pion-exchange model. Energy independent phase shift analyses of 
$\pi^+\pi^-$ scattering up to 1800 MeV have
been carried out
first by the CERN-Munich group 
\cite{Hyams:1973zf,Grayer:1974cr}
 (CM-I) using the assumptions of ``spin and phase coherence''
\cite{Ochs:1972mc}; results above 1400 MeV are superseded
by the more complete analyses based on CM-II data (see below). 
Above 1 GeV there are in general multiple phase shift
solutions which represent the same $\pi\pi$ angular distribution moments.
Such multiple solutions up to 1800 MeV 
have been obtained first by Estabrooks and Martin
\cite{Estabrooks:1974qd}. 
Based on an improved data
analysis the CERN-Munich group obtained a similar set of results with
some smaller errors \cite{Hyams:1975mc} (CM-II). A unique solution has been
found by combining with results 
from GAMS
Collaboration \cite{Alde:1998mc} on the $\pi^0\pi^0$ final state
\cite{Ochs:2006rb,Ochs:2013gi}. 
\begin{figure*}[t]
\includegraphics*[angle=0,width=4.0cm,bbllx=3.0cm,bblly=1.5cm,bburx=10.0cm,
bbury=8.2cm]{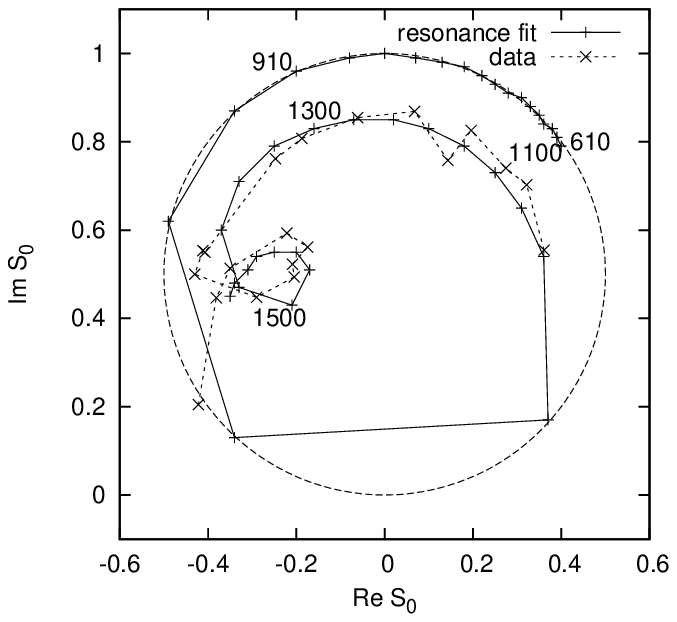} 
\includegraphics*[angle=0,width=4.0cm,bbllx=3.0cm,bblly=1.5cm,bburx=9.2cm,%
bbury=7.5cm]{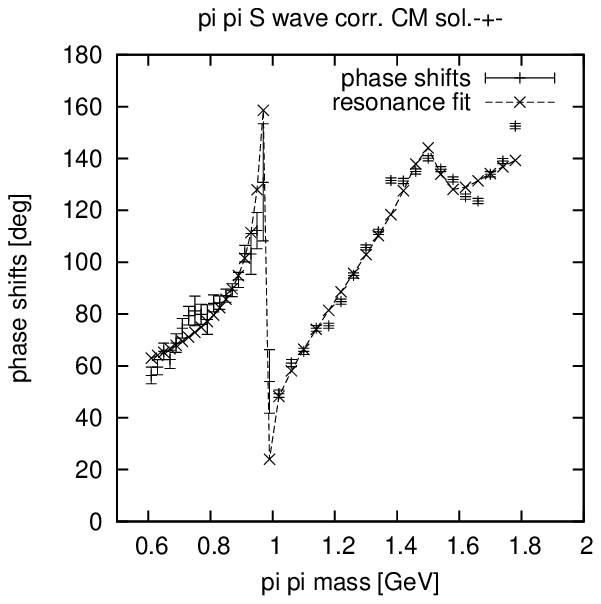} 
\includegraphics*[angle=0,width=4.0cm,bbllx=3.0cm,bblly=1.5cm,bburx=9.5cm,%
bbury=7.9cm]{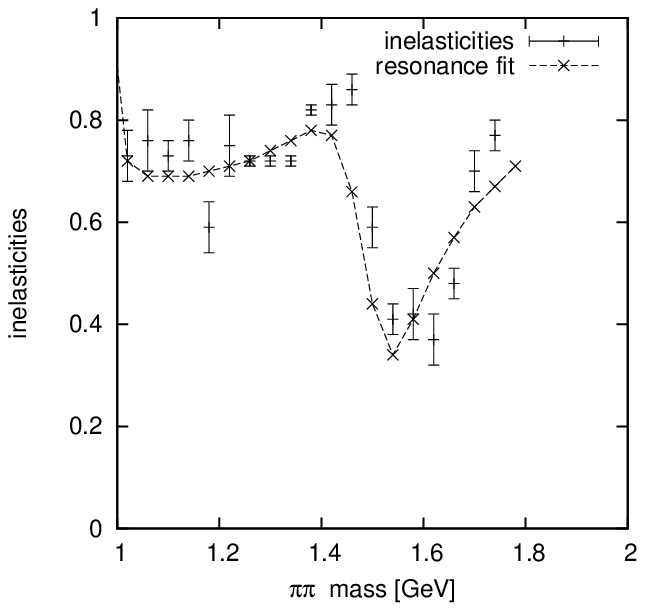} 
\caption{\label{fig:resonances} 
Data in $\pi\pi$ $S_0$ wave (CERN-Munich data CM-I/II):
Argand diagram for corrected $S_0$ wave,
phase shifts $\delta_0^0$ and inelasticities $\eta^0_0$; 
shown is also a preliminary resonance fit including
$f_0(500),\ f_0(980)$ and $f_0(1500)$.}
\end{figure*}
The isoscalar $S$ wave is shown in Fig.~\ref{fig:resonances}
where a clear signal  from
$f_0(1500)$ is seen: the resonance circle in the Argand diagram with
related movements of the phase and inelasticity near 1500 MeV. 
The elastic $\pi\pi$ width
is found as
\begin{equation}
f_0(1500): \quad  x_{\pi\pi}= 0.25\pm0.05 \ ({\rm CM-II}),
    \  x_{\pi\pi}=0.349 \pm 0.023\ ({\rm PDG}), \label{xpipi1500}
\end{equation}
where the first result (CM-II) is determined from Im $T_0$ of the
resonant elastic partial wave amplitude (from Fig.~\ref{fig:resonances}) 
and the second one (PDG) 
from all inelastic channel cross sections;  
both should agree because of the optical theorem and they roughly do within
30 \%.

There is no hint towards any resonance structure near 1370 MeV in any of
the plots of Fig.~\ref{fig:resonances} which leads to the limit
\begin{equation}
f_0(1370):\qquad  x_{\pi\pi} < 0.1 \quad [{\rm CL}=95\%] \quad
 {(\rm CM-II)}. \label{xpipi1370}  
\end{equation}
The absence of $f_0(1370)$ is
in agreement with the findings from an alternative phase shift analysis
\cite{Estabrooks:1978de}. 
On the other hand, 
global multi-resonance fits to the angular moment data (CM-I) with $f_0(1370)$
included have been presented in \cite{Bugg:2007ja} showing
an additional resonance circle. These results are in conflict with the
energy-independent bin-by-bin phase shift data in
Fig.~\ref{fig:resonances}. 

\subsection{Decays of $D$ and $B$ mesons}
In the weak decays of heavy quark mesons some well defined $q\bar q$
states evolve from the intermediate weak and strong interaction processes
and they finally can form isoscalar $f_0$ mesons. Recent results from $B$
factories and LHC have high statistical significance and they are well
suited to find small branching ratios.

\begin{figure}[t]
\includegraphics[angle=0,width=5.2cm]{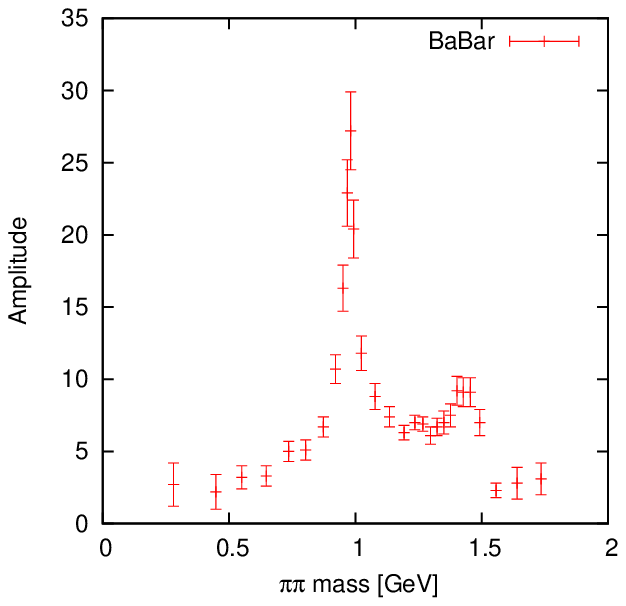} \hspace{-1.7cm}
\includegraphics[angle=0,width=5.2cm]{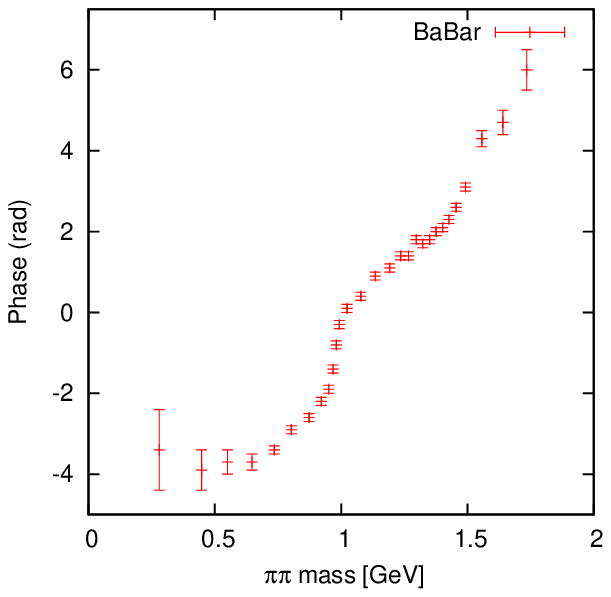} \hspace{-1.7cm}
\includegraphics[angle=0,width=5.2cm]{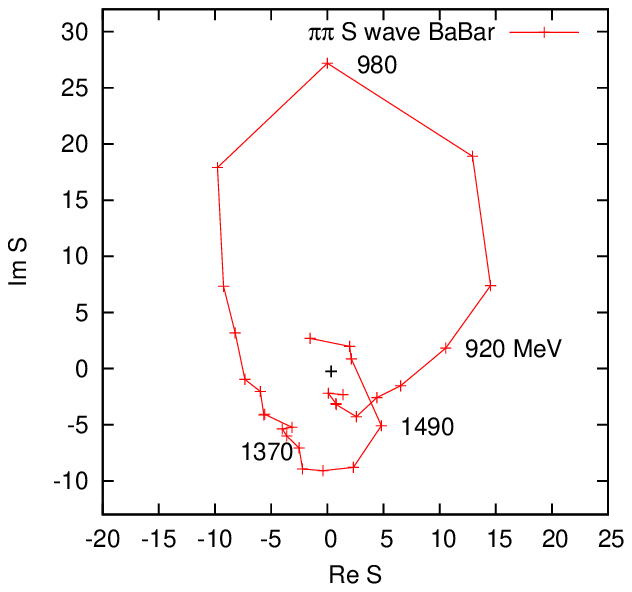}
\caption{\noindent $\pi\pi$ $S$ wave amplitude and phase extracted from
decays
$D_s^+\to \pi^+\pi^-\pi^+$ (BaBar Collaboration \cite{Aubert:2008ao});  
right panel: Argand diagram for $\pi\pi$ amplitude
(the phase is normalized to $\pi/2$ at the $f_0(980)$ peak). 
\label{fig:dsdec}}
\end{figure}

As an example, we report here  the decay $D_s^+\to \pi^+\pi^-\pi^+$ where the
dominant subprocess is identified as
$
D_s^+\to \pi^+ + s\bar s;\ s\bar s\to \pi^+\pi^-
$
with possible intermediate $f_0(1370)$ and $f_0(1500)$. An energy-independent 
phase shift analysis carried out by the BaBar Collaboration
\cite{Aubert:2008ao} is shown in Fig.~\ref{fig:dsdec}, see also
\cite{Ochs:2013gi}. One can see a strong movement of the amplitude related
to $f_0(980)$ and $f_0(1500)$, while there is no effect visible in between
where the phase movement becomes minimal.
A similar process is $B_s^0\to J/\psi+\pi^+\pi^-$ with 
subprocess $B_s^0\to J/\psi+s\bar s;\ s\bar s\to \pi^+\pi^-$ which has
been studied by the LHCb Collaboration  \cite{LHCb:2012ae}. Besides 
$f_0(980)$ one other resonance has been identified with parameters close to $f_0(1500)$
(see also \cite{Ochs:2013gi}).

\section{Scalar meson spectrum without $f_0(1370)$}
In view of the lacking evidence for $f_0(1370)$ alternative schemes for the
scalar spectrum have been looked for. In the approach by Minkowski and Ochs
\cite{Minkowski:1998mf} the lightest $q\bar q$ nonet includes
\begin{equation}
f_0(980),\ a_0(980),\ K^{*0}(1430),\ f_0(1500),
\end{equation} 
whereas the glueball is represented by $f_0(500)/\sigma$. It is assumed that
what is called $f_0(500)$ corresponds to the broad object centered at 1000
MeV with comparable width as observed in the $\pi\pi$ phase shift analysis of
Fig.~\ref{fig:resonances}: the phase shift passes $90^\circ$ near
1000 MeV after the effect from $f_0(980)$ is removed; so we call this state
also $f_0(500-1000)$. An object at this mass
is also found as lightest gluonic meson in the QCD sum rule approach
\cite{Narison:2005wc}. A similar $q\bar q$ nonet including $f_0(980)$ and
$f_0(1500)$ with a flavour mixing as for $\eta,\eta'$ but with reversed mass
ordering had been proposed before \cite{Klempt:1995ku,Dmitrasinovic:1996fi}.
No $K^{*0}(900)/\kappa$ is needed in this scheme; note that the phase movement
in $K\pi$ scattering related to $K^{*0}(900)/\kappa$ is only about
$40^\circ$ \cite{Ochs:2013gi}. 

Recently an attempt has been presented to determine 
the constituent structure of the light $f_0's$
from available branching fractions \cite{Ochs:2013gi}.
It is found that the $f_0(500)/\sigma$ decays 
are not ``flavour blind'' as expected for a glueball. The
flavour composition of $f_0(980)$ is found similar to the one of $\eta'$
confirming the earlier result \cite{Minkowski:1998mf}; the gluonic
component is estimated as $\lesssim 25\%$.
For $f_0(1500)$ a gluonic component is established as function of
the scalar mixing angle 
$\phi_{sc}$. Then, a minimal mixing scheme is proposed as
\begin{eqnarray}
|f_0(500-1000)\ran=& \sin \phi_G |q\bar q\ran - \cos\phi_G |gg\ran,\\ 
|f_0(1500)\ran=& \cos \phi_G |q\bar q\ran + \sin\phi_G |gg\ran;\
\end{eqnarray}
where $|q\bar q\ran =  \cos \phi_{sc} |n\bar n\ran - \sin\phi_{sc} 
       |s\bar s\ran$ is near a flavour octet with the mixing
 angle $\phi_{sc}=(30\pm3)^\circ$, and $\phi_G\sim 35^\circ$ in the
 simplest model. 
For  $\phi_G=0$ we recover the model \cite{Minkowski:1998mf}.
The strong mixing of the glueball into $f_0(500-1000)$ and $f_0(1500)$ 
is a feature also
found in recent QCD lattice calculations \cite{Hart:2006ps} and QCD sum rules
\cite{Harnett:2008cw}.    

\section{Concluding remarks}
After 40 years of experimental and theoretical work 
our knowledge on the scalar mesons has been considerably improved, but 
we have not yet succeded ultimately to proof the existence of the scalar
glueball and to determine its mass. 
The identification of a supernumerous state in the
nonet classification of mesons depends on the knowledge of all nonet members. 
This is difficult 
if broad objects like $f_0(1370)$ are involved
with small 2-body branching ratios if any. 

Therefore we have argued \cite{Ochs:2013gi} not to rely only on establishing
such difficult states like $f_0(1370)$ but to investigate other approaches
as well.

 1. Study of leading resonances in gluon jets (at large Feynman $x$).\\ 
Several LEP experiments observed a significant 
excess of neutral leading clusters beyond expectations from MC's as one 
expects from glueball production. The effect should be stronger at the LHC.

2. Study of decays of charmonium states like $\chi_c$ with primary $gg$
decay.\\ 
Pairs of scalar particles should be
produced according to flavour symmetry if they belong to the same $q\bar q$ 
multiplet, but deviations are expected for gluonic states.      

Such studies hopefully will provide new evidence for gluonic mesons if they
exist. 
\section*{Acknowledgement}
I would like to thank the organizers for providing the stimulating
athmosphere with useful discussions and to Peter Minkowski for the
collaboration.

\end{document}